\documentclass[11pt,a4paper]{article}
\usepackage[margin=2cm]{geometry}
\usepackage[utf8]{inputenc}
\usepackage[T1]{fontenc}
\usepackage{amsmath,amssymb,array,amsfonts}
\usepackage{booktabs}
\usepackage[sectionbib,round]{natbib}
\usepackage{graphicx}
\usepackage{fancyvrb}
\usepackage{alltt}
\RequirePackage[hyphens]{url}
\RequirePackage[pagebackref]{hyperref}

\RequirePackage{color}
\definecolor{link}{rgb}{0.45,0.51,0.67}
\hypersetup{
	colorlinks,%
	citecolor=link,%
	filecolor=link,%
	linkcolor=link,%
	urlcolor=link
}

\DeclareUnicodeCharacter{B1}{$\pm$}

\providecommand{\tightlist}{%
  \setlength{\itemsep}{0pt}\setlength{\parskip}{0pt}}

\newcommand{\CRANpkg}[1]{\href{https://CRAN.R-project.org/package=#1}{\pkg{#1}}}%

\let\pkg=\strong

\newcommand{\address}[1]{\addvspace{\baselineskip}\noindent\emph{#1}}

\DefineVerbatimEnvironment{Sinput}{Verbatim}{}
\DefineVerbatimEnvironment{Soutput}{Verbatim}{}
\DefineVerbatimEnvironment{Scode}{Verbatim}{}
\DefineVerbatimEnvironment{Sin}{Verbatim}{}
\DefineVerbatimEnvironment{Sout}{Verbatim}{}
\newenvironment{Schunk}{}{}

\makeatletter
\DeclareRobustCommand\code{\bgroup\@noligs\@codex}
\def\@codex#1{\texorpdfstring%
	{{\normalfont\ttfamily\hyphenchar\font=-1 #1}}%
	{#1}\egroup}
\providecommand{\operatorname}[1]{%
  \mathop{\operator@font#1}\nolimits}
\renewcommand{\P}{%
  \mathop{\operator@font I\hspace{-1.5pt}P\hspace{.13pt}}}
\newcommand{\E}{%
  \mathop{\operator@font I\hspace{-1.5pt}E\hspace{.13pt}}}

\makeatother

\begin{document}

\title{Binary R Packages for Linux:\\
Past, Present and Future}
\author{by Iñaki Ucar and Dirk Eddelbuettel}

\maketitle

\abstract{%
Pre-compiled binary packages provide a convenient way of efficiently
distributing software that has been adopted by most Linux package
management systems. However, the heterogeneity of the Linux ecosystem,
combined with the growing number of R extensions available, poses a
scalability problem. As a result, efforts to bring binary R packages to
Linux have been scattered, and lack a proper mechanism to fully
integrate them with R's package manager. This work reviews past and
present of binary distribution for Linux, and presents a path forward by
showcasing the `cran2copr' project, an RPM-based proof-of-concept
implementation of an automated scalable binary distribution system with
the capability of building, maintaining and distributing thousands of
packages, while providing a portable and extensible bridge to the system
package manager.
}

\newcommand{\inlinenote}[3]{{\color{#1} [#2: #3]\noindent\ }}
\newcommand{\IU}[1]{\inlinenote{olive}{IU}{#1}}
\newcommand{\DE}[1]{\inlinenote{teal}{DE}{#1}}

\vspace*{-1mm}

\hypertarget{introduction}{%
\section{Introduction}\label{introduction}}

\vspace*{-1mm}

R has become an essential tool, language, and environment for working
with data. One of the reasons for its success can be attributed to the
\emph{Comprehensive R Archive Network} (CRAN), a network of mirrored
sites \citep{RJ-2009-014}. It offers (at present) over 17,000 packages
in a way that ensures quality, interoperability, and, last but not
least, installability. R has also gone to great length to abstract away
the notion of the underlying operating system (OS). While users on a
particular OS may stress minor idiosyncratic aspects of the user
interface (such as clipboard copying between applications), both R code
and how users operate with R is essentially entirely portable between
systems.

Among the aforementioned 17,000 add-on packages extending R, one
particular commonality is the use of compiled code. This has several
advantages such as better performance for heavy computations as well as
possibly a tighter integration via the internals of R, but also direct
access to a wide variety of system and application libraries. To this
end, R provides official interfaces to Fortran, Java and C, and thus
effectively to any compiled language that can generate a C interface,
such as C++, Rust or Go. Currently, more than 24\% of CRAN packages use
compiled code, and another 57\% have at least one hard dependency
(\emph{Depends} or \emph{Imports}) that uses compiled code. In other
words, if we install a random package from CRAN, we have about a 81\%
chance of requiring compilation, directly or indirectly. Moreover, as
Fig. \ref{fig:needcomp} suggests, packages requiring direct compilation
are much more common among the most widely used packages: up to 50\% of
the most popular packages (according to the number of downloads)
directly use compiled code.

\begin{Schunk}
\begin{figure}

{\centering \includegraphics[width=0.85\linewidth]{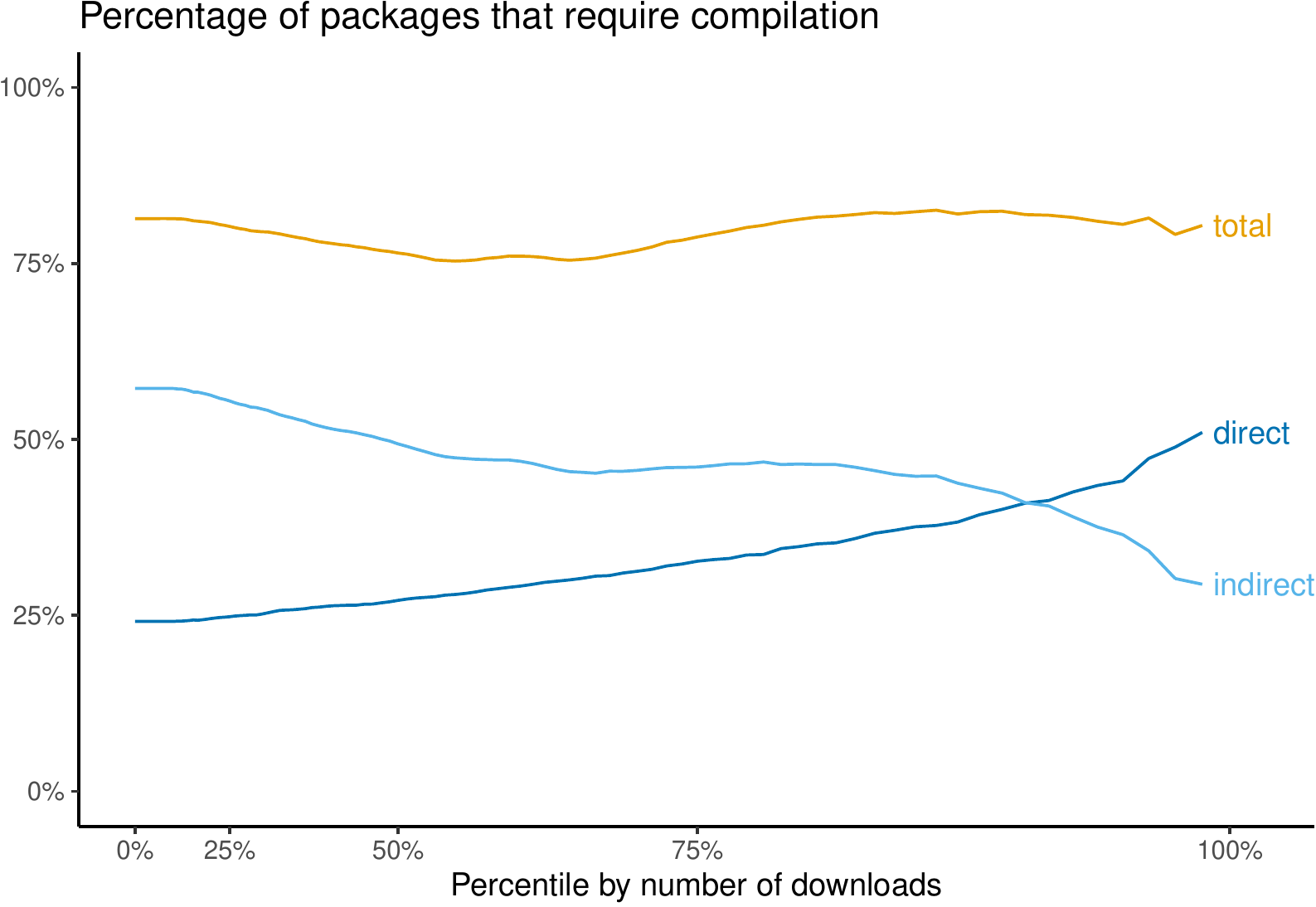} 

}

\caption[Percentage of packages that require direct or indirect compilation by popularity, measured by percentile of downloads from RStudio's CRAN logs for January 2021]{Percentage of packages that require direct or indirect compilation by popularity, measured by percentile of downloads from RStudio's CRAN logs for January 2021.}\label{fig:needcomp}
\end{figure}
\end{Schunk}

It should be noted that these performance advantages that make the
package attractive to the developer (and of course also to its users)
may come at a ``hidden'' yet high cost for the users. Installation of R
packages from source becomes more expensive (both in terms of time and
computational resources), and requires an appropriate toolchain to be
installed. Traditionally, Linux distributions have taken proper care of
this second aspect by providing a sufficient set of tools via the system
package manager, while Windows and macOS systems require extra steps and
add-ons. The availability of a single and relatively stable toolchain
and system libraries allowed CRAN to build the necessary infrastructure
to distribute binary packages for these systems, easing the burden on a
large portion of the user base. On the other hand, CRAN never provided
Linux binaries (except for some early experimental builds for Debian),
likely due to the fragmentation of Linux distributions leading to a
smaller market share, combined with the fact that Linux users have
historically been regarded as ``power'' users. As a consequence, Linux
users likely continue to spend considerable time and CPU cycles on
source installations.

Acknowledging that hard numbers are difficult to come by, anecdotal
evidence suggests that, still nowadays, the majority of users deploy
R---and write packages for R---on either Windows or macOS
\citep{SO:Survey:2020}. On the other hand, the Linux OS is very widely
deployed on ``cloud'' computing instances. And although Windows and
macOS are gaining better support on new continuous integration and
deployment (CI/CD) platforms, Linux proliferated sooner and faster, in
part thanks to its early support for OS-level virtualization
(\emph{i.e.}, container technologies such as Docker). It can then be
argued that portability and the widespread adoption of CI services had a
hand in an almost invisible usage increase on operating systems on which
users may not be working directly. As a result, we can conclude that
binary distribution for Linux would benefit a substantial portion of
both package installations and users that are not necessarily accustomed
to this platform. Although binary distribution is necessarily tied to a
specific toolchain---therefore imposing specific settings that might not
be according to the user's preference---it greatly helps data scientists
in getting whole groups of packages (such as the \emph{tidyverse}, or
the geospatial stack) quickly deployed in their systems.

\hypertarget{past-and-present}{%
\section{Past and present}\label{past-and-present}}

Binary package distribution faces a number of challenges. Here, we
identify and discuss four of them from the Linux perspective: OS
support, dependency management, scalability and integration.

\begin{description}
\tightlist
\item[OS support]
Cross-distribution incompatibilities can be a major source of concern
for binary packages. In the Linux ecosystem, this issue becomes more
dramatic as the matrix of possible configurations grows: there are
several families of distributions, supported by different package binary
formats; different distributions per family, with different philosophies
about naming conventions and update cycles; and, finally, there are
several supported versions per distribution at any given time too.
\item[Dependency management]
Once a package has been successfully built, it must be distributed along
with the same set of dependencies used for the build process. This is
not limited to other R packages, but it is particularly important for
code that links against external libraries and interfaces. No portable
solution for providing or querying system dependencies reliably exists.
Static linking and embedded libraries are popular ways to overcome both
OS support and dependency management issues, especially since the
development of packaging technologies such as Flatpak
\citep{flatpak:2021}, Snap \citep{snap:2021} or AppImage
\citep{appimage:2021}. However, these techniques may be better suited
for monolithic graphical applications, rather than for complex projects
such as the R ecosystem with its built-in extension mechanism.
\item[Scalability]
CRAN has experimented an impressive growth over its lifetime
\citep{JSSv073i02}. Building and distributing binary packages at scale
means handling the enormous complexity of versions and dependencies in
an automated way for thousands of packages.
\item[Integration]
CRAN's integrated support for Windows and macOS binaries entails a
friendly experience for the user, who can install thousands of packages
from the R command line using the built-in \texttt{install.packages}. On
Linux, however, package management has a different model in which
specialized system tools require administrative privileges and maintain
the integrity of the software stack.
\end{description}

CRAN itself hosted at least some Linux binaries at its very beginning
\citep{Hornik:CRAN:1997}. Since then, most Linux distributions provide
at least the R system and interpreter in their official repositories,
and may also maintain various subsets of add-on packages. This task is
extremely time-consuming for maintainers, because it typically requires
a per-package peer-review for the initial inclusion, and then the
application of a stringent set of policies and standards. Consequently,
these subsets are commonly very limited, and package versions tend to
lag behind.

In order to improve upon official repositories, several projects have
over time tried to tackle these challenges with varying degrees of
success. A short list of these attempts follows.

\begin{description}
\tightlist
\item[cran2deb (take one)]
Albrecht Gebhard wrote a first fully automated conversion of CRAN
packages into Debian binaries using Perl in the early 2000s, which lead
to subsequent work summarised in a \emph{useR!} presentation
\citep{useR:2007}.
\item[cran2deb (take two)]
Charles Blundell, as a Google Summer of Code student and during a
subsequent academic year, rewrote this in R with a SQL backend, which
one of us later ran for a while peaking at around 6k or 7k CRAN
packages---until it sadly broke while being presented at CRAN
headquarters in Vienna \citep{useR:2009}.
\item[debian-r]
Don Armstrong had a parallel setup during his PhD studies with up to 10k
packages, but then the RAID system broke \citep{debian-r:2015}.
\item[c2d4u]
During this time, Michael Rutter has (essentially singlehandedly) grown
a `cran2deb' offshoot on Ubuntu and its Launchpad build and hosting
system to about 4.7k packages across three releases
\citep{useR:2011, c2d4u:2021}.
\item[CRAN2OBS]
Independently of the work described here, a parallel (and long-running)
effort has focused on an automated system for turning CRAN packages into
RPM binaries for the SUSE Linux distribution at scale, with an
impressive CRAN coverage of over 16k packages as of August 2021
\citep{cran2obs:2020}. This initiative demonstrates the capabilities of
the openSUSE Build Service (OBS) \citep{obs:2021}, which is designed to
be cross-distribution. Hence, the successful demonstration of the use of
the OBS opens the door for future work which may consolidate efforts and
result in a unified service relying on a pan-distribution build service.
\end{description}

All these distribution-based solutions, either via official or
contributed repositories, inherently provide dependency management as
well as a tight integration with the OS. Another collateral advantage of
using system repositories is that package installations are shared
across users, so it is easier for a system administrator to maintain a
consistent and updated stack of packages. However, although `cran2deb'
and `CRAN2OBS' constitute a step forward towards scalable distribution,
the lack of integration constitutes a major drawback, especially for
casual Linux users.

Although the need for better integration of (binary) package
installations directly from R was also expressed occasionally
\citep{GilBellosta:Binaries:2010}, it was not until recently when new
projects started to focus on this challenge. Particularly, RStudio's
Public Package Manager \citep{rspm:2020}, based on the proprietary
product RSPM (RStudio Package Manager), stands out as the most
comprehensive solution to date. RSPM works by providing a drop-in
replacement for CRAN mirrors that serves Linux users pre-compiled
packages via \texttt{install.packages} (when available) with impressive
breadth of cross-platform coverage, yet no (formal) dependency support.
Thus, the Linux community benefits---under RStudio's terms of use---from
binary installations of CRAN packages, but many of such installations
may be fragile and break unless the user installs the required system
dependencies in a second (manual) step. And although RSPM alleviates the
task of discovering such dependencies for each supported distribution
via a dedicated (yet heuristic) database, the process of installing
system dependencies involves administrative rights and still remains a
separate workflow outside R, which ultimately hinders full integration.

\hypertarget{cran2copr-rpm-distribution-of-cran-packages-at-scale}{%
\section{cran2copr: RPM distribution of CRAN packages at
scale}\label{cran2copr-rpm-distribution-of-cran-packages-at-scale}}

In this context, the `cran2copr' project \citep{cran2copr:2020} was
created as an initiative under the umbrella of the Fedora Project that
aims at distributing all packages on CRAN via a separate RPM repository
\citep{iucar/cran:2020}, with full dependency management, automatic
daily synchronization and seamless integration with the R console. It
consists of a Copr project for building and distributing packages; the
\pkg{CoprManager} package for the integration, and standalone R tooling
around the Copr API for automation. Building on the experience of
previous projects, `cran2copr' shares the advantages of
distribution-based solutions and solves the remaining major challenges
of binary distribution for Linux: scalability and integration (see Table
\ref{tab:projects} for a comparison).

\begin{Schunk}
\begin{table}

\caption{\label{tab:projects}Main challenges of binary package distribution and degree of achievement for different projects. A checkmark denotes fulfillment, while a bullet indicates partial fulfillment.}
\centering
\begin{tabular}[t]{rccccr}
\toprule
 & Official repos & cran2deb & CRAN2OBS & RSPM & cran2copr\\
\midrule
OS support & $\bullet$ & $\bullet$ & \checkmark & \checkmark & $\bullet$\\
Dependencies & \checkmark & \checkmark & \checkmark &  & \checkmark\\
Scalability &  & $\bullet$ & \checkmark & \checkmark & \checkmark\\
Integration &  &  &  & $\bullet$ & \checkmark\\
\bottomrule
\end{tabular}
\end{table}

\end{Schunk}

\hypertarget{a-primer}{%
\subsection{A primer}\label{a-primer}}

This section briefly demonstrates the user workflow. The following
commands enable the project's Copr repository, and install the
\pkg{CoprManager} utility on Fedora 32:

\begin{verbatim}
$ sudo dnf install 'dnf-command(copr)'
$ sudo dnf copr enable iucar/cran
$ sudo dnf install R-CoprManager
\end{verbatim}

From this point on, no further administrative tasks are required.
Package installations are carried out through R's standard
\texttt{install.packages} function, which transparently delegates in the
system package manager. In the event that a certain R package is not
available in the Copr repository, \pkg{CoprManager} also provides an
automatic fallback mechanism, and \texttt{install.packages} continues
the installation directly from CRAN.

This is how the installation looks like for a package included in the
project (\CRANpkg{units}) and another that is currently not
(\CRANpkg{gifski}). As can be seen, the request for \CRANpkg{units} is
transparently intercepted and installed in binary form from the Copr
repository, including its system dependencies; \CRANpkg{gifski}, on the
other hand, is pulled directly from CRAN in standard source form.

\begin{Schunk}
\begin{Sinput}
install.packages(c("gifski", "units"))
\end{Sinput}
\begin{Soutput}
#> Install system packages as root...
#> (1/3): R-CRAN-units-0.6.7-3.fc32.x86_64.rpm    5.2 MB/s | 787 kB     00:00
#> (2/3): R-CRAN-Rcpp-1.0.5-2.fc32.x86_64.rpm     2.0 MB/s | 2.0 MB     00:00
#> (3/3): udunits2-2.2.26-6.fc32.x86_64.rpm       420 kB/s | 617 kB     00:01
#>   Preparing        :                                                        1/1
#>   Installing       : udunits2-2.2.26-6.fc32.x86_64                          1/3
#>   Installing       : R-CRAN-Rcpp-1.0.5-2.fc32.x86_64                        2/3
#>   Installing       : R-CRAN-units-0.6.7-3.fc32.x86_64                       3/3
#>   Running scriptlet: R-CRAN-units-0.6.7-3.fc32.x86_64                       3/3
#>   Verifying        : R-CRAN-Rcpp-1.0.5-2.fc32.x86_64                        1/3
#>   Verifying        : R-CRAN-units-0.6.7-3.fc32.x86_64                       2/3
#>   Verifying        : udunits2-2.2.26-6.fc32.x86_64                          3/3
#> 
#> Installing package into ‘/home/user/R/x86_64-redhat-linux-gnu-library/4.0’
#> (as ‘lib’ is unspecified)
#> trying URL 'https://cloud.r-project.org/src/contrib/gifski_0.8.6.tar.gz'
#> [...]
#> 
#> * DONE (gifski)
\end{Soutput}
\end{Schunk}

There is no integration with \texttt{remove.packages} by default, which
will only remove packages installed in the user library. Removing system
packages is still possible through the \pkg{CoprManager} API:

\begin{Schunk}
\begin{Sinput}
CoprManager::remove_copr("units")
\end{Sinput}
\begin{Soutput}
#> Remove system packages as root...
#>   Preparing        :                                                        1/1
#>   Erasing          : R-CRAN-units-0.6.7-3.fc32.x86_64                       1/3
#>   Erasing          : R-CRAN-Rcpp-1.0.5-2.fc32.x86_64                        2/3
#>   Erasing          : udunits2-2.2.26-6.fc32.x86_64                          3/3
#>   Running scriptlet: udunits2-2.2.26-6.fc32.x86_64                          3/3
#>   Verifying        : R-CRAN-Rcpp-1.0.5-2.fc32.x86_64                        1/3
#>   Verifying        : R-CRAN-units-0.6.7-3.fc32.x86_64                       2/3
#>   Verifying        : udunits2-2.2.26-6.fc32.x86_64                          3/3
\end{Soutput}
\end{Schunk}

As another advantage, it can be noted that the system package manager
automatically removes dependencies that are not required by any other
package. Finally, the integration can be disabled, so that
\texttt{install.packages} only installs from CRAN:

\begin{Schunk}
\begin{Sinput}
CoprManager::disable()
install.packages("units")
\end{Sinput}
\begin{Soutput}
#> Installing package into ‘/home/user/R/x86_64-redhat-linux-gnu-library/4.0’
#> (as ‘lib’ is unspecified)
#> trying URL 'https://cloud.r-project.org/src/contrib/Rcpp_1.0.5.tar.gz'
#> trying URL 'https://cloud.r-project.org/src/contrib/units_0.6-7.tar.gz'
#> [...]
\end{Soutput}
\end{Schunk}

Only that now the installation from source fails, because the required
\texttt{udunits2} library is not available.

\hypertarget{building-and-maintaining-thousands-of-binary-packages}{%
\subsection{Building and maintaining thousands of binary
packages}\label{building-and-maintaining-thousands-of-binary-packages}}

The `cran2copr' project is based on Fedora Copr, a platform to build and
manage contributed package repositories based on the Copr Buildsystem
\citep{copr:2020}. Copr projects publish standard RPM repositories that
are readily available through \texttt{dnf}'s \texttt{copr} command, as
shown in the previous section. The build system supports several
RPM-based distributions and a wide range of architectures (either
natively or through virtualization). As a result, Copr projects are able
to build and distribute binary packages for several versions of a number
of distributions (currently, Fedora, CentOS, Mageia, openSUSE,
OpenMandriva and Oracle Linux) and multiple architectures (currently,
i386, x86\_64, aarch64, armhfp, ppc64le and s390x).

This project leverages the Copr infrastructure to manage thousands of
packages and builds with a set of standalone R scripts around the Copr
API. The workflow for building a package is as follows. A single-file
SPEC (a ``recipe'' for building an RPM package) is auto-generated from a
standard template, the package's DESCRIPTION file, and a list of (non-R)
system dependencies (see the next subsection for further details). Then,
it is uploaded to the Copr Buildsystem via its API, and no further local
processing is required. Note that SPECs are committed to the project's
repository for the purposes of caching and keeping a history of changes,
but this step is not required.

Using this SPEC, the Copr Buildsystem automatically fetches the sources
of the package to build the intermediate source RPM, and then spawns
build tasks for all the configured chroots (combinations of Linux
distribution, version and architecture). Builds are always carried out
in a fresh isolated environment consisting on a minimal set of packages.
This strategy, along with the automatic execution of the standard set of
RPM checks, has proved to be useful to detect a whole new class of
installation issues that go unnoticed on CRAN. \footnote{Just to mention
  some of them, by examining build failures, we have found undeclared
  dependencies, incorrect versioning of dependencies, and a number of
  security issues, such as pre- and post-installation downloads,
  incorrect execution flag in files, or the inclusion of the buildroot
  path in installed files. This has led to many upstream reports to
  package maintainers as wells as CRAN maintainers and R developers.}
Once binary packages are built, they are automatically signed and
included in the exported RPM repository for each chroot.

In contrast to previous open projects, `cran2copr' is completely
stateless and requires no complex setup, making it very easy to automate
via CI/CD services. In fact, we take advantage of GitHub Actions
(GitHub's integrated CI system) to perform updates on a schedule:
everyday at 00:00 UTC, the CRAN database is fetched and compared against
the Copr repository to remove archived packages and submit updated SPECs
for building.

Another key challenge to maintaining thousands of binary packages is the
need to perform mass rebuilds due to breaking changes (\emph{e.g.}, in a
new version of R, or in system libraries). The Copr Buildsystem exposes
a batching functionality that makes this process straightforward. First,
a simple iterative algorithm takes the CRAN database and derives a list
of lists of packages, so that each batch only requires packages from the
previous batch. Second, builds for each package are requested via the
Copr API, indicating the batch identifier and the dependency on the
previous batch. Typically, a single mass rebuild is required for each
Fedora release, and it is performed in the rawhide branch after the
official mass rebuild is carried out per release schedule.

Finally, to be able to scale to CRAN's current size, a number of design
decisions were made to accommodate some limitations of traditional
packaging \citep{fedora-packaging-guidelines:2021}:

\begin{itemize}
\tightlist
\item
  Package names, versions and license strings are copied from
  DESCRIPTION files unmodified, and files under the installation
  directory are not explicitly listed.
\item
  Packages are built as-is, without unbundling any included third-party
  libraries.
\item
  To ensure compatibility with the official repositories, the
  \texttt{/usr/local} prefix is used for package installation. Thus, all
  R packages are installed under \texttt{/usr/local/lib/R/library}.
\end{itemize}

Currently, `cran2copr' maintains almost the whole CRAN (over 17.5k
packages as of August 2021), except for a few hundred packages (out of
about 17,000) due to various reasons. For example, Bioconductor
\citep{gentleman2004bioconductor} is not supported at present. Thus,
CRAN packages that require (\emph{Depends} or \emph{Imports})
Bioconductor packages are excluded. Packages that require unsupported
system dependencies are excluded too. Several cases fall under this
category, such as packages that depend on proprietary libraries
available on CRAN, or packages that require an Internet connection at
installation time. For instance, this is the case with \CRANpkg{gifski}
shown in the previous section, as well as with other Rust-based
packages. These are excluded because enabling repository-wide Internet
connection is a security risk, and Copr still lacks support for this
option on a per-package basis.

Despite these limitations, as Copr---which is under heavy
development---grows in capabilities and resources, `cran2copr' will be
able to improve even further to bring binary distribution of R packages
to more distributions and architectures.

\vspace*{-1mm}

\hypertarget{a-note-about-systemrequirements}{%
\subsection{\texorpdfstring{A note about
\texttt{SystemRequirements}}{A note about SystemRequirements}}\label{a-note-about-systemrequirements}}

As established by the \emph{Writing R Extensions} manual
\citep{R-exts:2021}, ``dependencies external to the R system should be
listed in the \texttt{SystemRequirements} field''. This usually includes
the need for specific libraries (such as the \CRANpkg{units} package in
our example above, which requires the \texttt{udunits2} library) as well
as programs (such as a compiler, \texttt{pandoc}, specific
\LaTeX~classes\ldots), and it is therefore an essential part of the SPEC
generation process. However, the complete lack of standardization of
this field, together with the fact that sometimes packages are accepted
on CRAN with undeclared dependencies, makes manual intervention
unavoidable.

There have been previous attempts to formalizing these requirements,
\footnote{The second `cran2deb' version, `debian-r' and `c2d4u' all
  use(d) a local SQLite database. R-Hub maintans a repository at
  \url{https://github.com/r-hub/sysreqsdb}, as does RStudio's system
  requirements for R packages at
  \url{https://github.com/rstudio/r-system-requirements}.} but we opted
for maintaining our separate database of system requirements to better
distinguish between build- and run-time requirements, thus minimizing
the footprint for the builder as well as for the final user. This
database was initially bootstrapped by scrapping the
\texttt{SystemRequirements} field from the DESCRIPTION file for all
packages on CRAN, and curated by hand. Maintenance consists in examining
build failures from time to time, and manually updating the database
accordingly.

Although some automation may be possible by parsing the
\texttt{SystemRequirements} field with some clever regular expression
(to our knowledge, `CRAN2OBS' takes this approach), such methods are
doomed to failure in the long run. Any long-term solution to this
problem would probably involve the commitment of CRAN to help
standardize, enforce and maintain such a database of system
requirements.

\vspace*{-1mm}

\hypertarget{integrating-rootless-binary-installation-on-linux}{%
\subsection{Integrating rootless binary installation on
Linux}\label{integrating-rootless-binary-installation-on-linux}}

The last challenge that previous projects always lacked was a
transparent integration with the R console. Recently, RStudio's RSPM
system partially achieved this goal, but at a cost: a complete
decoupling from the system repositories. The unified package management
on Linux, central to automatic dependency management, seems to
irrevocably require specific tools and administrative privileges.

To overcome this issue, \pkg{CoprManager} integrates a system service
written in Python that exposes a D-Bus interface. The D-Bus
\citep{dbus:2020}, or \emph{Desktop Bus}, is a convenient and mature
abstraction that is omnipresent in modern Linux environments. It
provides inter-process communication (IPC) through a remote procedure
call (RPC) mechanism handled by the D-Bus daemon, which starts and stops
registered services and manages messages for them.

When \pkg{CoprManager} is installed and enabled, it traces the
\texttt{install.packages} function and forwards the requested packages
over this interface. The system service receives the names of the R
packages and translates them to the proper names in the system
repositories (\emph{e.g.}, \texttt{units} to \texttt{R-CRAN-units}).
With the necessary privileges to talk to the system package manager, it
requests the installation and reports the progress to the R session.
Finally, if a package is not found, it is returned to the unprivileged R
session, so that \texttt{install.packages} can continue the installation
from CRAN (or other repositories).

This mechanism bridges the gap between robust package management through
system repositories and the user-friendliness of R's
\texttt{install.packages}. Moreover, it can be easily extended to
support fine-grained permission management through D-Bus policies or
Polkit to, \emph{e.g.}, enable just a subset of users or packages.

\hypertarget{bridge-to-system-package-manager}{%
\section{\texorpdfstring{\CRANpkg{bspm}: Bridge to System Package
Manager}{: Bridge to System Package Manager}}\label{bridge-to-system-package-manager}}

The experience with `cran2copr' allowed us to distill the main features
of \pkg{CoprManager} into the \CRANpkg{bspm} package \citep{cran:bspm}.
This is an extensible cross-distribution solution that provides other
projects, such as `cran2deb', with the previously missing integration
layer. In fact, the most recent versions of \pkg{CoprManager} are built
from \CRANpkg{bspm}`s sources as a ``branded'' version specifically
targeted to the 'cran2copr' project. Moreover, \CRANpkg{bspm} implements
further enhancements that enable the same integration in other major use
cases beyond desktop environments. Already existing examples for both
Debian and Ubuntu uses of \CRANpkg{bspm} are working with containers,
CI/CD workflows, and direct use via a \CRANpkg{littler}
\citep{cran:littler} wrapper.

\hypertarget{extensible-cross-distribution-support}{%
\subsection{Extensible cross-distribution
support}\label{extensible-cross-distribution-support}}

Cross-distribution support is achieved by exposing a standardized
interface in the underlying Python service. This interface comprises
three calls:

\begin{itemize}
\tightlist
\item
  \texttt{discover}: Receives no input, and explores the system
  repositories to find the list of prefixes required to map R package
  names to system package names (\emph{e.g.}, \CRANpkg{units} to
  \texttt{r-cran-units} for Ubuntu/Debian) via a set of heuristics. This
  mapping---as well as an optional exclusion list---can also be preset
  by an administrator to enforce a particular subset of packages.
\item
  \texttt{install} / \texttt{remove}: Both receive a list of R package
  names, and try to install/remove them from the system repositories
  using the previous mapping. Errors from non-existent packages are
  caught, and those R package names are returned to the R session. This
  provides a fallback mechanism within the \texttt{install.packages}
  integration, which transparently continues the installation from
  source.
\end{itemize}

D-Bus-mediated calls to the Python service also report the identifier
(PID) of the calling process, so that progress can be injected into the
R session via its standard error file descriptor. These three functions
are prototypes and need to be implemented for each specific system
package manager. Currently, \CRANpkg{bspm} implements connectors to the
two major package managers, \texttt{dnf} (for the \texttt{.rpm}-based
distributions, and \texttt{apt} for the \texttt{.deb}-based ones),
covering a large portion of the Linux ecosystem.

\hypertarget{beyond-the-desktop-environment}{%
\subsection{Beyond the desktop
environment}\label{beyond-the-desktop-environment}}

The \CRANpkg{bspm} package was conceived with the primary goal of
providing full integration with the R console in a desktop environment
setup, as well as in multi-tenant server deployments. In these
environments, \CRANpkg{bspm} provides unprivileged R sessions with a
secure and standardized mechanism to request binary package
installations to a privileged service. However, there are other
important use cases in which the D-Bus interface is not available.
Instead, cloud deployments, as well as CI/CD systems, generally consist
of minified headless containers based on Docker or a similar
virtualization technology, either running as root or as an unprivileged
user with access to \texttt{sudo}. In such scenarios, \CRANpkg{bspm}
works transparently by directly calling the bridge interface without the
need for an intermediate service.

\CRANpkg{bspm} is used by container images by the Rocker Project
\citep{RJ-2017-065}, a widely-used suite of Docker images with
customized R environments for particular tasks. Specifically, the
\texttt{r-bspm} images leverages a simple setup via a
\texttt{r-cran-bspm} binary package for either Debian (using the
`testing' distribution) or Ubuntu (using the two most recent LTS
releases). The `r-ci' project \citep{r-ci:2021} utilises this setup with
its \CRANpkg{bspm}-driven delivery of pre-build binaries under full
dependency control in a portable script that provides flexible and
lightweight CI use at GitHub Actions, Travis, and Azure Pipelines.

\hypertarget{summary-and-future-work}{%
\section{Summary and future work}\label{summary-and-future-work}}

This paper reviews the history of the distribution of binary R packages
for Linux and presents `cran2copr', an RPM-based proof-of-concept
implementation of an automated scalable binary distribution system. Our
solution not only demonstrates the feasibility of building, maintaining
and distributing thousands of binary packages---potentially for multiple
Linux distributions and multiple architectures---but also provides a
portable and extensible bridge to the system package manager, thus
providing both full dependency resolution and integration with the R
console via \CRANpkg{bspm}.

The huge success of the RSPM repositories currently kindly provided for
free by RStudio to the R community is one of the most clear indicators
of the importance of binary package distribution for Linux. This work
makes the case for future community-driven cross-distribution
developments based on modern open-source build systems like Copr,
discussed here, or the more general OBS. Build systems present several
advantages to current CRAN infrastructure, such as isolation and
scaling, which would enhance the quality of packages by exposing them to
more stringent checks as well as a wider and more realistic set of
environments. But more importantly, artifacts from the building process
would be automatically gathered into a consistent (possibly,
multi-architecture) set of contributed repositories that would benefit
the R community as a whole.

\bibliography{cran2copr}

\address{%
Iñaki Ucar\\
Universidad Carlos III de Madrid\\%
UC3M-Santander Big Data Institute\\ Calle Madrid 135\\ 28903 Getafe
(Madrid), Spain\\
ORCiD
\href{https://orcid.org/0000-0001-6403-5550}{0000-0001-6403-5550}\\%
\href{mailto:inaki.ucar@uc3m.es}{\nolinkurl{inaki.ucar@uc3m.es}}%
}

\address{%
Dirk Eddelbuettel\\
University of Illinois at Urbana-Champaign\\%
Department of Statistics\\ Illini Hall, 725 S Wright St\\ Champaign, IL
61820\\ ORCiD
\href{https://orcid.org/0000-0001-6419-907X}{0000-0001-6419-907X}\\
\url{https://dirk.eddelbuettel.com}\\%
\href{mailto:dirk@eddelbuettel.com}{\nolinkurl{dirk@eddelbuettel.com}}%
}

\end{document}